\documentclass[showpacs,showkeys,preprint,nofootinbib]{revtex4}

\usepackage{amsxtra}

\usepackage{amsmath, amsthm, amsfonts, amssymb, bbm}
\usepackage[mathscr]{eucal}

\theoremstyle{plain}
\newtheorem{proposition}{Proposition}
\newtheorem{theorem}[proposition]{Theorem}
\newtheorem{lemma}[proposition]{Lemma}

\theoremstyle{definition}

\theoremstyle{remark}
\newtheorem{remark}{Remark}

\newcommand{\ud}{\mathrm{d}}
\newcommand{\del}{\partial}

\newcommand{\Tr}{\mathrm{Tr}}

\newcommand{\V}[1]{\mathbf{#1}}

\newcommand{\schw}{\mathcal{S}}

\newcommand{\R}{\mathbb{R}}
\newcommand{\supp}{\mathrm{supp}}

\newcommand{\betrag}[1]{\left| #1 \right|}
\newcommand{\norm}[1]{\left\| #1 \right\|}

\newcommand{\pv}[1]{\mathrm{P} \frac{1}{#1}}
\newcommand{\iep}[1]{\frac{1}{#1+i\epsilon}}
\newcommand{\iem}[1]{\frac{1}{#1-i\epsilon}}

\newcommand\zpiw[1]{\frac 1 {{\sqrt{2 \pi}}^{#1}}}
\newcommand\zpi[1]{\frac 1 {{(2 \pi)}^{#1}}}

\begin{document}

\title{Infrared cutoffs and the adiabatic limit in noncommutative spacetime}
\author{Claus D\"oscher}
\email{claus.doescher@desy.de}
\author{Jochen Zahn}
\email{jochen.zahn@desy.de}
\affiliation{ II.~Institut f\"ur Theoretische Physik, Universit\"at Hamburg, Luruper Chaussee 149, 22761 Hamburg, Germany}
\affiliation{Zentrum f\"ur Mathematische Physik, Universit\"at Hamburg, Bundesstra\ss e 55, 20146 Hamburg, Germany}

\preprint{DESY 05-251}
\preprint{ZMP-HH/05-24}

\begin{abstract}
We discuss appropriate infrared cutoffs and their adiabatic limit for field theories on the noncommutative
Minkowski space in the Yang-Feldman formalism. In order to do this, we consider a mass term as interaction term.
We show that an infrared cutoff can be defined quite analogously to the commutative case and that the adiabatic
limit of the two-point function exists and coincides with the expectation, to all orders.
\end{abstract}

\maketitle

\section{Introduction}

Already since the 1930s there are speculations that spacetime has a noncommutative structure at small scales. A
model that is particularly interesting, motivated from the discussion of limitations of optimal localization of
experiments \cite{DFR} and string theory in a constant background $B$-field \cite{Schomerus}, is the
noncommutative Minkowski space. It is generated by selfadjoint coordinates $q^{\mu}$ that fulfill the
commutation relations
\begin{equation*}
  [q^{\mu}, q^{\nu}] = i \sigma^{\mu \nu}.
\end{equation*}
For the purpose of our study, $\sigma$ can be any antisymmetric matrix, i.e., we also allow for so-called
space-time noncommutativity ($\sigma^{0i} \neq 0$).

There are several inequivalent approaches to quantum field theory on the noncommutative Minkowski space which
all have some advantages and disadvantages (for a discussion, see e.g. \cite{Bahns}). The modified Feynman rules
proposed by Filk \cite{Filk} are easy to handle from a computational point of view. However, they lead to a
non-unitary $S$-matrix in the case of space-time noncommutativity \cite{Gomis}. Nevertheless, they are widely
used and in this context the issue of UV/IR-mixing was first discussed \cite{Minwalla}: Nonplanar graphs become
UV-finite due to phase factors, but this regularization disappears when the external momentum goes to zero. This
raises the question of the renormalizability of such models.

The Hamiltonian approach first proposed by Doplicher, Fredenhagen and Roberts \cite{DFR} leads (with a suitable definition of the product of quantum fields) to a UV-finite scalar field theory \cite{UVfinite}. However, there are unresolved IR-problems. Furthermore, the free and the interaction part of the Hamiltonian are treated in different ways, which is quite unnatural in gauge theories. In fact, there are problems with the Ward identities already at the tree level~\cite{Ohl}.

In the case of space-time noncommutativity, the most promising approach is the Yang-Feldman formalism
\cite{YangFeldman}, which has been proposed for the study of noncommutative field theories by Bahns et al
\cite{BDFP}. However, this formalism is underdeveloped in comparison to the other approaches to quantum field
theory, also in the commutative case. There are open questions, on the conceptual as well as on the
computational level. On the conceptual side, there is the issue of an appropriate infrared cutoff that makes the
perturbative expansion well-defined. Of course, we finally ,i.e., after computation of the $n$-point function that interests us,
want to remove this cutoff, i.e., carry out the adiabatic limit.
Since mass terms have in general the worst infrared behavior, we will focus on a mass term as interaction term
in a massive scalar field theory and compute the two-point function of the interacting field. On the computational side,
this will help to identify mass and field-strength renormalizations in interacting models.


Furthermore, it is a crucial feature of perturbation theory that the parameters of a given model are determined
by renormalization conditions and do not have to coincide with the "bare" parameters that are used to set up the
theory. In particular, one may add a finite mass term $L_{int} = - \mu \phi^2$ in order to account for a
physical mass that is not the free or bare mass $m$. For the consistency of the theory it is crucial that this
is equivalent to defining $\sqrt{m^2+\mu}$ as the free mass\footnote{In the context of algebraic perturbation
theory, this requirement was formulated as a renormalization condition by Hollands and Wald \cite{HollandsWald}.
They call it the principle of perturbative agreement.}. Of course, this has to be understood in the sense of
formal power series in $\mu$.

We start with a discussion of the problem in the commutative case. In order to avoid infrared problems, we multiply the interaction mass term with a test function $g$. Our goal is to compute the adiabatic limit ($g \to 1$) of the two-point function
\begin{equation}
\label{eq:2ptfh}
  \langle \phi(f) \phi(h) \rangle
\end{equation}
of the interacting field, which, in the Yang-Feldman formalism\footnote{In the present case this is of course
equivalent to the definition via retarded products (see, e.g., \cite{DuetschFredenhagen}).}, is defined as a
formal power series $\phi = \sum_{n=0}^{\infty} \mu^n \phi_n$ that fulfills the equation of motion
\begin{equation*}
  ( \Box + m^2 ) \phi = - \mu g \phi.
\end{equation*}
Obviously, $\phi_0$ satisfies the free field equation. We identify it with the incoming field. Then the higher
terms are recursively defined by
\begin{equation*}
  \phi_n(x) = - \int \ud^4y \ \Delta_{R}(x-y) g(y) \phi_{n-1}(y).
\end{equation*}
Here $\Delta_R$ is the retarded propagator. At $n$th order in $\mu$, the two-point function~(\ref{eq:2ptfh}) is then
\begin{equation}
\label{eq:2ptn}
  \sum_{k=0}^n \langle \phi_k(f) \phi_{n-k}(h) \rangle.
\end{equation}
A well-known theorem of Epstein and Glaser \cite{EpsteinGlaser} guarantees the existence of the adiabatic limit.
For the convenience of the reader we included it in Appendix \ref{app:EG}. Of course, we expect to find
\begin{equation}
\label{eq:2ptnlimit}
  (2\pi)^2 \int \ud^4k \ \check{f}(k) \check{h}(-k) \frac{1}{n!} \left( \frac{\ud}{\ud m^2} \right)^n \hat{\Delta}_+^{m^2}(k)
\end{equation}
in the adiabatic limit. That this is indeed true is the statement of the following
\begin{theorem}
\label{MainTheorem} Let $\{ \check{g}_a \}_a$ be a sequence of Schwartz function with support in a closed subset
of $V_n=\left\{ p \in \R^4 | \betrag{p^0} < 2 m/n \right\}$ that converges to $(2 \pi)^2 \delta^{4}$ in the
sense of rapidly decreasing distributions. Then the adiabatic limit ($a \to \infty$) of (\ref{eq:2ptn}) is
(\ref{eq:2ptnlimit}).
\end{theorem}
We will prove this theorem in Sec.~\ref{sec:2ptlimit}.

Going to the noncommutative case, there are two immediate problems.
The first is to find an  appropriate infrared cutoff, i.e., the analog of multiplying with $g$ as above. Then
the next question is the existence of the adiabatic limit. Of course, there is no Epstein-Glaser theorem for the
noncommutative Minkowski space. The phase factors that are responsible for the UV/IR-mixing mentioned above may
spoil the calculation. It is thus not clear if the adiabatic limit exists at all, if it is unique (i.e.,
independent of the sequence of ``cutoff functions''), and if it yields the expected result. If this was not the case, we would have difficulties to make any sense of these theories. We discuss these questions in Sec.~\ref{sec:NCFT}.
We show that an infrared cutoff can be defined quite analogously to the commutative case and that it yields the correct adiabatic limit.
This paves the way for a forthcoming study of dispersion relations in interacting noncommutative field theories in the Yang-Feldman formalism~\cite{InPrep}.

Our conventions are summarized in Appendix~\ref{app:conventions}. Appendix~\ref{app:EG} contains the theorem of Epstein and Glaser mentioned above. In Appendix~\ref{app:lemma} we prove a technical lemma.

\section{The commutative case}
\label{sec:2ptlimit}

We start with proving Theorem \ref{MainTheorem} for the case $n=1$. In Sec.~\ref{subsec:n} we do this for all
orders.

\subsection{First order}

For $n=1$, (\ref{eq:2ptn}) is given by
\begin{align*}
  & - \int \ud^4x_0 \ud^4x_1 \ud^4x_2 \ f(x_0) h(x_2) g_a(x_1) \\
  & \qquad \left[ \Delta_R(x_0-x_1) \Delta_+(x_1-x_2) + \Delta_+(x_0-x_1) \Delta_A(x_1-x_2) \right] \\
  = & - (2\pi)^2\int \ud^4k_0 \ud^4k_1 \ \check{f}(k_0) \check{h}(-k_1) \check{g}_a(k_1-k_0) \left[ \hat{\Delta}_R(k_0) \hat{\Delta}_+(k_1) + \hat{\Delta}_+(k_0) \hat{\Delta}_A(k_1) \right].
\end{align*}
Using~(\ref{eq:Delta_R})
and setting $\pm x = k^0_{0/1} - \omega_{0/1}$, this can be written as
\begin{multline}
\label{eq:1st_order}
  \frac 1{2\pi}\int \frac{\ud^3 k_0}{2 \omega_0} \frac{\ud^3 k_1}{2 \omega_1} \ud x \ \check{g}_a(\omega_1 - \omega_0 -x, \V{k_1} - \V{k_0}) \\
  \left[ \check{f}(\omega_0+x, \V{k_0}) \check{h}(-\omega_1, -\V{k_1}) \left( \iep{x} - \iep{x + 2 \omega_0} \right) \right. \\
  \left. - \check{f}(\omega_0,  \V{k_0}) \check{h}(-\omega_1+x, -\V{k_1}) \left( \iep{x} - \iep{x - 2 \omega_1} \right) \right].
\end{multline}
We first deal with the two terms involving $\iep{x}$. We write
\begin{align*}
  \check{f}(\omega_0+x,  \V{k_0}) & = \check{f}(\omega_0,  \V{k_0}) + x F(\omega_0+x,  \V{k_0}) \\
  \check{h}(-\omega_1+x, -\V{k_1}) & = \check{h}(-\omega_1, -\V{k_1}) + x H(-\omega_1+x, -\V{k_1})
\end{align*}
where $F$ and $H$ are smooth functions with $F(\omega_0,  \V{k_0}) = \del^0 \check{f}(\omega_0,  \V{k_0})$ and $H(-\omega_1, -\V{k_1}) = \del^0 \check{h}(-\omega_1, -\V{k_1})$. Then the terms of zeroth order in $x$ cancel.  The terms of first
order in $x$ now yield
\begin{equation*}
  2\pi\int \ud^3k \ \frac{1}{(2\omega_{\V k})^2}
           \left[ \del_0 \check{f}(\omega_\V{k}, \V{k}) \check{h}(-\omega_{\V k}, - \V{k})
           - \check{f}(\omega_\V{k}, \V{k}) \del_0 \check{h}(-\omega_{\V k}, - \V{k}) \right]
\end{equation*}
in the adiabatic limit. 
\begin{remark}
Note that it was crucial here to consider the sum of $\langle \phi_0(f) \phi_1(h) \rangle$ and $\langle
\phi_1(f) \phi_0(h) \rangle$. The individual terms are divergent. This is a nice illustration of Remark 4 in
\cite{EpsteinGlaser}.
\end{remark}
It remains to treat the two terms involving $\iep{x \pm 2 \omega_{0/1}}$. If we assume that $\check{g}_a$ is
supported in a closed subset of $V_1 = \left\{ p \in \R^4 | \betrag{p^0} < 2 m \right\}$, then the singularity
$x = \mp 2 \omega_{0/1}$ lies outside the support of $\check{g}_a$. Thus, we may carry out the adiabatic limit
and obtain
\begin{equation*}
  -{2\pi}\int \ud^3k \ \frac{1}{(2\omega_{\V k})^3}
           \left[ \check{f}(\omega_\V{k}, \V{k}) \check{h}(-\omega_{\V k}, - \V{k})
           - \check{f}(\omega_\V{k}, \V{k}) \check{h}(-\omega_{\V k}, - \V{k}) \right].
\end{equation*}
Combining all this, we get
\begin{multline*}
 2\pi\int \ud^3k \ \left( - \frac{1}{4 \omega_\V{k}^3} \check{f}(\omega_\V{k},\V{k}) \check{h}(-\omega_\V{k},-\V{k}) \right. \\
  \left. + \frac{1}{4 \omega_\V{k}^2} \left[ \del_0 \check{f}(\omega_\V{k},\V{k}) \check{h}(-\omega_\V{k},-\V{k})
    - \check{f}(\omega_\V{k},\V{k}) \del_0 \check{h}(-\omega_\V{k},-\V{k}) \right] \right).
\end{multline*}

Thus, the adiabatic limit of (\ref{eq:2ptn}) exists for $n=1$. We still have to check that it coincides with (\ref{eq:2ptnlimit}). We compute
\begin{align*}
 \frac{\ud}{\ud m^2} \Delta_+^{m^2}(f,h)  = &- (2\pi)^2 \int \ud^4k \ \check{f}(k) \check{h}(-k) \frac 1 {2\pi}\theta(k_0) \delta'(k^2-m^2) \\
 = &- 2\pi \int \ud^3k \int_0^{\infty} \frac{\ud s}{2 \sqrt{s}} \check{f}(\sqrt{s}, \V{k}) \check{h}(-\sqrt{s}, -\V{k}) \delta'(s-\omega_{\V{k}}^2) \\
 = & 2\pi\int \ud^3k \ \left( -\frac{1}{4 \omega_\V{k}^3} \check{f}(\omega_\V{k}, \V{k}) \check{h}(-\omega_\V{k}, -\V{k}) \right. \\
    & \ \left. + \frac{1}{4 \omega_\V{k}^2} \left[ \del_0 \check{f}(\omega_\V{k}, \V{k}) \check{h}(-\omega_\V{k}, -\V{k})
      - \check{f}(\omega_\V{k}, \V{k}) \del_0 \check{h}(-\omega_\V{k}, -\V{k}) \right] \right).
\end{align*}
Thus, we have proven Theorem \ref{MainTheorem} for $n=1$.

\subsection{$n$th order}
\label{subsec:n}
At $n$th order, the two-point function is
\begin{multline*}
 (-1)^n \int \prod_{i=0}^{n+1} \ud^4y_i \ f(y_0) h(y_{n+1}) \prod_{i=1}^n g_a(y_i) \\
  \sum_{m=0}^n \Delta_R(y_0-y_1) \dots \Delta_R(y_{m-1} -y_m) \Delta_+(y_m-y_{m+1}) \Delta_A(y_{m+1}-y_{m+2}) \dots \Delta_A(y_n -
  y_{n+1}).
\end{multline*}
In momentum space this is
\begin{multline*}
 (-1)^n (2\pi)^2\int \prod_{i=0}^{n} \ud^4k_i \ \check{f}(k_0) \check{h}(-k_{n}) \prod_{l=1}^{n} \check{g}_a(k_{l} - k_{l-1}) \\
  \sum_{m=0}^{n} \hat{\Delta}_R(k_0) \dots \hat{\Delta}_R(k_{m-1}) \hat{\Delta}_+(k_m) \hat{\Delta}_A(k_{m+1}) \dots
  \hat{\Delta}_A(k_{n}).
\end{multline*}
Thus, in the notation of Appendix \ref{app:EG}, we have
\begin{multline}
\label{eq:F}
 \hat{F}_R(p,q) = (-1)^n (2\pi)^2 \int \prod_{i=0}^{n} \ud^4k_i \ \delta(p_1-k_0) \delta(p_2+k_n)  \prod_{l=1}^{n} \delta(q_l-k_{l}+ k_{l-1}) \\
  \sum_{m=0}^{n} \hat{\Delta}_R(k_0) \dots \hat{\Delta}_R(k_{m-1}) \hat{\Delta}_+(k_m) \hat{\Delta}_A(k_{m+1}) \dots \hat{\Delta}_A(k_{n}).
\end{multline}

We discuss the support properties of $F_R$. By definition, it has retarded support (see Appendix \ref{app:EG}).
By using the advanced instead of the retarded propagator, one gets the distribution $F_A$. It is obtained from
$F_R$ by interchanging the retarded and advanced propagators in (\ref{eq:F}). With a slight modification of the
argument of Epstein and Glaser (see Appendix \ref{app:EG}) one can show that the difference
$\hat{F}_R(p,q)-\hat{F}_A(p,q)$ vanishes for $q \in R_n$ given by~(\ref{eq:R}).
\begin{remark}
In the present case, this can be shown in a direct way: Write
\begin{equation*}
  \hat{\Delta}_{R/A}(k) = \hat{\bar{\Delta}}(k) \pm \frac{1}{2} \hat{\Delta}(k) =  \zpi{2}\left(-\pv{k^2-m^2} \pm i \pi \epsilon(k^0) \delta(k^2-m^2)\right).
\end{equation*}
If one now computes the difference $\hat{F}_R-\hat{F}_A$, all the terms with an even number of $\hat{\Delta}$s
drop out and we are left with
\begin{multline*}
  \frac{(-1)^n}{(2\pi)^{2n-1}} \sum_{\substack{I \subset \{1, \dots, n\} \\ \betrag{I} \text{ even}}} (i \pi)^{\betrag{I}-1} \prod_{i \in I} \delta(k_i^2-m^2) \prod_{j \in I^c} \text{P} \frac{-1}{k_j^2-m^2} \\
  \sum_{m \in I} \left( \prod_{\substack{l \in I \\ l < m}} \epsilon(k_l^0) \right) \theta(k_m^0) \left( \prod_{\substack{l \in I \\ l>m}} \epsilon(-k_l^0) \right).
\end{multline*}
Due to the fact that $\betrag{I}$ is even, the sum over $m$ will vanish if all the $\{k_i^0\}_{i \in I}$ are positive. Now we assume that the support of the $\check{g}$s is such that their product vanishes if $(k_i - k_j)^2 < 4 m^2 \ \forall i,j$. Then the terms where some of the $\{k_i^0\}_{i\in I}$ are negative vanish because $(k_i -k_j)^2
\geq 4 m^2$ when $k_i$ and $k_j$ are on opposite mass shells.
\end{remark}
Thus, we can conclude from Theorem~\ref{thm:EG} that $\hat{F}_R(p,q)$ is a tempered distribution in $p$ and infinitely differentiable in $q$ for $q \in R_n$. Thus, it is possible to integrate it (over $q$) with a rapidly decreasing distribution with support in $R_n$. Note that if $\check{g}$ has support in $V_n$, the product of the $\check{g}$s will have support in $R_n$.


Using (\ref{eq:Delta_R}) we can write the above as
\begin{multline*}
  \zpi{2n-1}\int \prod_{i=0}^{n} \frac{\ud^4k_i}{2\omega_i}  \check{f}( k_0) \check{h}(-k_{n}) \prod_{l=1}^n \check{g}_a( k_{l} - k_{l-1}) \\
  \sum_{m=0}^{n} \left[ \delta(k_m^0-\omega_m) \prod_{j=0}^{m-1} \left( \iep{k_j^0 - \omega_j} - \iep{k_j^0 + \omega_j} \right)\right. \\
   \left.\prod_{j=m+1}^{n} \left( \iem{k_j^0 - \omega_j} - \iem{k_j^0 + \omega_j} \right) \right].
\end{multline*}
We assume that the $\check g_a$s have support in a closed subset of  $V_n$. Then
$\frac{1}{k_j^0 + \omega_j \pm i \epsilon}$ can not become singular, so we may write
\begin{equation*}
   \frac{1}{k_j^0 - \omega_j \pm i \epsilon} - \frac{1}{k_j^0 + \omega_j \pm i \epsilon} = \frac{1}{k_j^0 - \omega_j \pm i \epsilon} \frac{2 \omega_{j}}{k_j^0 + \omega_j}.
\end{equation*}
We abbreviate
\begin{equation*}
  T(k) = \frac{2 \omega_\V{k}}{k^0 + \omega_\V{k}}.
\end{equation*}

Now we Taylor expand $\check{f}$, $\check{h}$ and $T$ in $x_j = k_j^0 - \omega_j$, i.e. we write
\begin{equation*}
  \check{f}(\omega+x,\V{k}) = \sum_{k=0}^{n} \frac{x^k}{k!} \del_0^k \check{f}(\omega, \V{k}) + \frac{x^{n+1}}{(n+1)!} \tilde{f}(\omega+x, \V{k})
\end{equation*}
where $\tilde{f}$ is smooth.
$\check{h}(-k)$ and $T(k)$ are expanded similarly.

The strategy is now as follows: We show that the term involving only the polynomial parts of the Taylor expansion is finite and gives exactly the desired result. Then it remains to show that the terms that involve at least one rest term of the Taylor expansion vanish. This will be done by a scaling argument that refers to the theorem of Epstein and Glaser.

We start with the first task. With the coordinate transformation
\begin{gather*}
  -y_1 = x_1 - x_0; \quad \dots \quad -y_{m-1} = x_{m-1} - x_{m-2} ; \quad -y_{m} =  - x_{m-1} \\
  -y_{m+1} = x_{m+1} ; \quad - y_{m+2} = x_{m+2} - x_{m+1} ; \quad  \dots \quad  - y_n = x_n - x_{n-1}
\end{gather*}
and using $\del_0^l T(\omega_k, \V{k}) = l! (-2 \omega_k)^{-l}$, we may write the term involving only the polynomial parts as
\begin{multline*}
  \zpi{2n-1}\int \prod_{i=0}^{n} \frac{\ud^3k_i}{2\omega_i} \prod_{j=1}^{n}  \ud y_j \
  \prod_{l=1}^n \check{g}_a( \omega_{l} - \omega_{l-1} - y_l, \V{k}_{l} - \V{k}_{l-1}) \\
  \sum_{m=0}^{n} \Biggl[ (-1)^{n-m} \prod_{s=1}^{m} \iep{\sum_{t=s}^{m} y_t}   \prod_{r=m+1}^n \iep{\sum_{t=m+1}^{r} y_t}  \\
  \left( \sum_{l_0=0}^{n} \frac{\left(- \sum_{t=1}^{m} y_t \right)^{l_0}}{l_0!} (-\del_0)^{l_0}\check{f}(\omega_0,\V{k}_0) \right) \prod_{s=1}^{m} \left( \sum_{l_{s}=0}^{n} \frac{\left(- \sum_{t=s}^{m} y_t \right)^{l_{s}}}{\left(2\omega_{s-1} \right)^{l_{s}}} \right) \\
  \prod_{r=m+1}^{n} \left( \sum_{l_{r+1}=0}^{n} \frac{\left( \sum_{t=m+1}^{r} y_t \right)^{l_{r+1}}}{\left(2\omega_{r}\right)^{l_{r+1}}} \right) \left( \sum_{l_{n+1}=0}^{n} \frac{\left(\sum_{t=m+1}^{n} y_t \right)^{l_{n+1}}}{l_{n+1}!} \del_0^{l_{n+1}} \check{h}(-\omega_{n}, -\V{k}_{n}) \right) \Biggr].
\end{multline*}
Now we need the following lemma, which is proven in Appendix~\ref{app:lemma}:
\begin{lemma}
\label{lemma:iep}
\begin{equation*}
  \sum_{m=0}^n (-1)^m \prod_{r=0}^{m-1}\frac{(-\sum_{t=r+1}^m y_t)^{l_r}}{\sum_{t=r+1}^m y_t+i\epsilon}
  \cdot \delta_0^{l_m}\cdot \prod_{r=m+1}^{n}\frac{(\sum_{t=m+1}^r y_t)^{l_r}}{\sum_{t=m+1}^r y_t+i \epsilon}
  =P_{l_0,\ldots,l_n}(y_1,\ldots,y_n),
\end{equation*}
where $P_{l_0,\ldots,l_n}$ is a homogeneous polynomial of degree $d:=\sum_{r=0}^n l_r-n$ if $d\ge 0$ and equals
 $1$ for $d=0$ and $0$ for $d<0$.
\end{lemma}

Thus, 
we get
\begin{equation*}
  2\pi\sum_{\sum_{t=0}^{n+1} l_t = n} \int \ud^3k \ \frac{(-1)^{n+l_{0}}}{(2 \omega_\V{k})^{n+1+\sum_{t=1}^{n} l_t}} \frac{1}{l_0!} \frac{1}{l_{n+1}!} \del_0^{l_0} \check{f}(\omega_\V{k}, \V{k}) \del_0^{l_{n+1}} \check{h}(-\omega_\V{k}, -\V{k})
\end{equation*}
in the adiabatic limit. This is
\begin{equation*}
   2\pi\sum_{l_1+l_2 \leq n} \int \ud^3k \ \frac{c_{n,l_1+l_2}}{(2 \omega_\V{k})^{2n+1-l_1-l_2}}
   (-1)^{l_1} {l_1+l_2 \choose l_2}  \del_0^{l_1} \check{f}(\omega_\V{k}, \V{k}) \del_0^{l_2} \check{h}(-\omega_\V{k}, -\V{k})
\end{equation*}
with
\begin{equation*}
  c_{n l} = \frac{(-1)^{n}}{l!} \frac{1}{2^{2n+1-l}} { 2n-l \choose n}.
\end{equation*}
It is straightforward to show that this coincides with (\ref{eq:2ptnlimit}).

It remains to show that the terms involving $\tilde{f}$, $\tilde{h}$ or $\tilde{T}$ vanish in the adiabatic
limit. We sketch how to do this.
Since Epstein and Glaser have shown that the adiabatic limit is independent of the sequence of test functions, we may restrict our attention to a limited class. We will assume factorization, i.e., $\check{g}_a(k) = \check{g}_a^t(k^0) \check{g}_a^s(\V{k})$, compact support, and scaling behavior: $\supp \check{g}_a \subset B(C a^{-1})$, $\norm{\check{g}_a^t}_{\alpha} < C'_\alpha a^{1+\betrag{\alpha}}$, and $\norm{\check{g}_a^s}_{\alpha} < C''_\alpha a^{3+\betrag{\alpha}}$. Here we used the notation $B(z) = \{ k \in \R^4 | \betrag{k} <z \}$ and $\norm{f}_{\alpha} = \sup_k \betrag{\del^{\alpha} f}$. Now, from $\betrag{\omega_1 - \omega_2} \leq \betrag{\V{k_1}-\V{k_2}}$ it follows that
\begin{equation*}
  \prod_{l=1}^n \check{g}_a^t(\omega_l - \omega_{l-1} + x_l - x_{l-1}) \check{g}_a^s(\V{k_l}-\V{k_{l-1}}) |_{x_m=0}
\end{equation*}
has support in $x_l \in [- c a^{-1}, c a^{-1}] \ \forall l$ for some constant $c$. Furthermore, we have
\begin{equation*}
  \sup_{y \in [-c a^{-1}, c a^{-1}]} \betrag{ \del^{\alpha}_y \int \ud x \ \frac{x^n}{x\pm i \epsilon} g_a(y-x)} < c'_\alpha a^{1-n+\betrag{\alpha}}
\end{equation*}
for some constant $c'_\alpha$. It follows that after carrying out the $n$ integrations over the $x$s, we get a smooth bounded function of the $\omega$s whose bound decreases at least as $a^{-1}$. Then we may carry out the $\V{k}$ integrations and end up with a sequence that decreases at least as $a^{-1}$. This concludes the proof of Theorem \ref{MainTheorem}.

\begin{remark}
\label{rem:support}
So far we have shown that the right adiabatic limit in $n$th order is obtained if the functions $\check
g_a$ have support in $V_n$. This seems to be rather restrictive, as for example $\check g_a$ cannot be analytic
and therefore $g_a$ cannot have compact support. But the result can be generalized to functions which can be
decomposed into
\begin{equation*}
  g_a=g^0_a+g^1_a
\end{equation*}
with $g^0_a$ as before and $\check g^1_a$ having arbitrary support but going to $0$ fast enough in $a$, that is,
\begin{equation*}
 \check g^1_a(k_1) \cdot \prod_{t=2}^n \check g^0_a(k_t)
  \xrightarrow[a \to \infty]{} 0 \text{ in } \schw(\R^{4 n}).
\end{equation*}
An example for such a sequence $g_a$ is easily constructed if we take an arbitrary function $g \in \schw(\R^4)$
with $g(0)=1$ and scale it, i.e., $g_a(x):=g(x/a)$. For the decomposition we then take a cutoff function $b \in
\schw(\R^4)$ with $b(k)=0$ if $\betrag{k^0} \geq 2 m/n$ and $b(k)=1$ if $\betrag{k^0} <  m/n$ and define $\check
g^0_a(k)= b(k) \check g_a(k)$, $\check g^1_a(k)= (1-b(k)) \check g_a(k)$. Thus, also cutoff functions with
compact support in position space can be used.
\end{remark}

\begin{remark}
So far we considered the two-point function obtained by taking the vacuum expectation value. Since the
construction of the interacting field is purely algebraic, one might wonder if one can use some other
(quasifree) state. This, however, is not possible. First of all, the theorem of Epstein and Glaser is not
applicable in this case. A direct computation for the case $n=1$ shows that if one uses a scaling test function
sequence one obtains a bounded sequence. Thus, there are cluster points, but most probably no unique limit that
is independent of the choice of the test function sequence.
\end{remark}

\section{The noncommutative case}
\label{sec:NCFT}



In order to prepare our analysis, we introduce some notation. Following \cite{DFR}, the free field on the
noncommutative Minkowski space $\mathcal{E}_{\sigma}$ is defined by
\begin{equation*}
  \phi(q) = (2\pi)^{-2} \int \ud^4k \ \hat{\phi}(k) \otimes e^{-ikq}.
\end{equation*}
Here $\hat{\phi}(k)$ is the usual field operator on the Fock space. In order to get a well-defined operator one
has to apply some suitable state on the $\mathcal{E}_{\sigma}$-part of the tensor product. Such states are
easily constructed if one introduced the trace on $\mathcal{E}_{\sigma}$: $\Tr e^{ikq} = (2 \pi)^2 \delta^4(k)$.
This is the analog of the integral in the commutative case. Suitable states are now given by $\phi(f) = \Tr \phi
f$, where $f$ is a test function. By this we mean
\begin{equation*}
  f(q) = \frac{1}{(2 \pi)^2} \int \ud^4k \ \check{f}(k) e^{i k q},
\end{equation*}
with $\check{f} \in \schw(\R^4)$.

Now the Yang-Feldman equation for a mass term is formally given by
\begin{equation}
\label{eq:phi_n_q}
  \phi_n(q) = \int \ud^4x \ \Delta_R(x) \phi_{n-1}(q-x).
\end{equation}
However, the question of an appropriate infrared cutoff and its adiabatic limit has not been discussed in great detail\footnote{In \cite{Bahns} it was proposed to multiply the retarded propagator in (\ref{eq:phi_n_q}) with a test function $g(x)$, compute the relevant $n$-point functions, and then take the adiabatic limit $g \to 1$. However, one can show that even on the ordinary (commutative) Minkowski space one does not obtain the correct adiabatic limit, already at first order.}.


In \cite{Zahn} the following type of infrared cutoff was proposed. Using the trace introduced above, one can define the quadratic interaction term
\begin{equation}
\label{eq:int_nc}
  S_{int} = - \mu \Tr \left( g^1 \phi g^2 \phi \right).
\end{equation}
Here $g^1$ and/or $g^2$ are test functions. This is analogous to the procedure in the commutative case.
Obviously, if the algebra was commutative, this would be the same as using $g = g^1 g^2$ as cutoff function.
Then the equation of motion is
\begin{equation*}
  \left( \Box + m^2 \right) \phi = - \mu/2  \left( g^1 \phi g^2 + g^2 \phi g^1 \right).
\end{equation*}
Note that for the perturbation series to be well-defined it is not important that both $g^1$ and $g^2$ are test functions. In fact, we might set $g^2 = \mathbbm 1$ from the beginning.

At $m$th order, the interacting field is\footnote{Here we use the Weyl formula $e^{ik_\mu q^\mu}e^{il_\nu
q^\nu}=e^{-i k_\mu \sigma^{\mu\nu} l_\nu/2}e^{i(k+l)_\mu q^\mu}$.}
\begin{multline*}
  \phi_m(q) = \frac{(-1)^m}{(2\pi)^{2(m+1)}} \int \ud^4p \prod_{j=1}^m (\ud^4k_j \ud^4 l_j) \ \hat{\phi}_0(p) e^{i(-p+ \sum_{t=1}^m (k_t+l_t)) q}  \\
   \prod_{r=1}^m \left[ \check{g}^1(k_r) \check{g}^2(l_r) \hat{\Delta}_R(p- \sum_{t=1}^r (k_t+l_t))
   \cos\left(\frac 12\bigl\{(k_r-l_r)\sigma \bigl[-p+\sum_{t=1}^{r-1} (k_t+l_t)\bigr]+k_r \sigma l_r\bigr\}\right)\right].
\end{multline*}
Thus, at $n$th order, the two-point function is
\begin{multline*}
  \frac{(-1)^n}{(2\pi)^{2(n-1)}} \int \prod_{j=1}^{n} (\ud^4k_j \ud^4l_j) \ud^4p \ud^4\tilde p \ud^4p_1 \ud^4p_2\
  \check{f}(p_1) \check{h}(p_2) \prod_{j=1}^n \check{g}^1(k_j) \check{g}^2(l_j) \hat{\Delta}_+(p) \delta(p+\tilde p)\\
  \sum_{m=0}^n \Biggl[ \delta(p_1 -p+ \sum_{t=1}^m (k_t+l_t))\ \delta(p_2-\tilde p+ \sum_{t=m+1}^n (k_t+l_t)) \\
  \prod_{r=1}^m \hat{\Delta}_R( p - \sum_{t=r}^m (k_t+l_t))\
    \cos\left(\frac 1 2 \bigl\{(k_r-l_r)\sigma\bigr[-p+\sum_{t=r+1}^m(k_t+l_t)\bigr]+k_r\sigma l_r\bigr\}\right)\\
  \prod_{r=m+1}^n \hat{\Delta}_R( \tilde p - \sum_{t=m+1}^r (k_t+l_t))\
    \cos\left(\frac 1 2 \bigl\{(k_r-l_r)\sigma\bigl[-\tilde p+\sum_{t=m+1}^{r-1}(k_t+l_t)\bigr]+k_r\sigma
    l_r\bigr\}\right)\Biggr],
\end{multline*}
where we have relabeled in each summand the variables $k_j$ and $l_j$ from 1 to m:
\begin{equation*}
  k_1 \to k_m , \ k_2\to k_{m-1} \ldots \qquad \text{and} \qquad l_1 \to l_m , \ l_2\to l_{m-1} \ldots ,
\end{equation*}
and use the higher indices for the parts coming from $\phi(h)$. Performing the $\tilde p$ and the $p$ integration, we obtain
\begin{multline*}
  \frac{(-1)^n}{(2\pi)^{2(n-1)}} \int \prod_{j=1}^{n} (\ud^4k_j \ud^4l_j) \ud^4p_1 \ud^4p_2 \
  \check{f}(p_1) \check{h}(p_2) \prod_{j=1}^n \check{g}^1(k_j) \check{g}^2(l_j) \\
  \sum_{m=0}^n  \Bigg[ \hat{\Delta}_+(p_1 + \sum_{t=1}^m (k_t+l_t))\ \delta(p_1+p_2+ \sum_{t=1}^n (k_t+l_t)) \\
  \prod_{r=1}^m \hat{\Delta}_R( p_1 + \sum_{t=1}^{r-1} (k_t+l_t))\
    \cos\left(\frac 1 2 \bigl\{(k_r-l_r)\sigma\bigl[-p_1-\sum_{t=1}^r(k_t+l_t)\bigr]+k_r\sigma l_r\bigr\}\right)\\
  \prod_{r=m+1}^n \hat{\Delta}_R( -p_1 - \sum_{t=1}^r (k_t+l_t))\
    \cos\left(\frac 1 2 \bigl\{(k_r-l_r)\sigma\bigl[p_1+\sum_{t=1}^{r-1}(k_t+l_t)\bigr]+k_r\sigma l_r\bigr\}\right)
    \Bigg].
\end{multline*}
In order to extend the sum in the argument of the second $\cos$ to $r$, we have to subtract $2k_r \sigma l_r$. As $\cos$ is even these factors can now be pulled outside the sum, and we obtain
\begin{multline*}
  \zpi {2n}\int \prod_{j=1}^{n} (\ud^4k_j \ud^4l_j) \ud^4p_1 \ud^4p_2 \
  \check{f}(p_1) \check{h}(p_2) \prod_{j=1}^n \check{g}^1(k_j) \check{g}^2(l_j) \\
  \prod_{r=1}^n \cos\left(\frac 1 2 \bigl\{(k_r-l_r)\sigma\bigl[p_1+\sum_{t=1}^r(k_t+l_t)\bigr]-k_r\sigma
  l_r\bigr\}\right)\\
 \hat{F}_R(p_1, p_2, k_1+l_1,\ldots,k_n+l_n).
\end{multline*}
The last two lines are again a tempered distribution in the $p$s and infinitely differentiable in the $k$s and $l$s as long as each $k_j+l_j$ lies inside $V_n$. In order to achieve this, we may for example require $\check{g}^1$ and $\check{g}^2$ to have support in a closed subset of $V_{2n}$. Obviously, the adiabatic limit exists and since the cosines give $1$ there, it is the usual one. Thus, we have proven the following
\begin{theorem}
Let $\phi$ be the interacting field defined via the Yang-Feldman formalism for the interaction term~(\ref{eq:int_nc}) with $g^i$ replaced by sequences $\{g_a^i\}_a$.
Let $\{\check{g}^1_a\}_a$ and $\{\check{g}^2_a\}_a$ be sequences of Schwartz functions with support in a closed subset of $V_{2n}$ that both converge to $(2\pi)^2 \delta^4$ in the sense of rapidly decreasing distributions. Or let $\check{g}^2_a = (2\pi)^2 \delta^4$ and $\{\check{g}^1_a\}_a$ be a sequence of Schwartz functions with support in a closed subset of $V_{n}$ that converges to $(2\pi)^2 \delta^4$ in the sense of rapidly decreasing distributions.
In both cases the abiabatic limit of (\ref{eq:2ptn}) is 
(\ref{eq:2ptnlimit}).
\end{theorem}
\begin{remark}
As was already mentioned in Remark~\ref{rem:support}, it is possible to allow also for test functions $\check{g}$ with larger support.
\end{remark}


\appendix

\section{Conventions and useful formulae}
\label{app:conventions}
We use the signature $(+,-,-,-)$. $\schw$ denotes the Schwartz space, $\schw'$ its dual.
The Fourier transformation and its inverse in $d$ dimensions are defined by
\begin{equation*}
  \hat f (k) =\zpiw{d}\int \ud^d x \ f(x) e^{i k x}; \qquad \check f(k)=\hat f (-k).
\end{equation*}
Free field of mass $m$:
\begin{equation*}
  \hat{\phi}(k) = \sqrt{2 \pi} \delta(k^2-m^2) \left[ \theta(k_0) a(\V{k}) + \theta(-k_0) a^{\dagger}(\V{-k}) \right].
\end{equation*}
Propagators:
\begin{align}
  \Delta_+(x-y)& =\langle\phi(x)\phi(y)\rangle \nonumber \\
  \hat \Delta_+(p)& =\frac 1 {2 \pi} \theta (p_0) \delta(p^2-m^2)
  =\frac 1 {2 \pi} \frac {\delta(p_0-\omega_\V p)}{2 \omega_\V p} \nonumber \\
  \hat \Delta_R(p)&=\lim_{\epsilon \searrow 0}\zpi{2}\frac {-1}{p^2-m^2+i \epsilon p_0} \nonumber \\
\label{eq:Delta_R}
  &=\lim_{\epsilon \searrow 0}\zpi{2} \frac 1 {2\omega_\V p}
  \left(\frac 1 {p_0+\omega_\V p+i \epsilon}-\frac 1 {p_0-\omega_\V p+i \epsilon}\right) \\
  \Delta_A(x)&=\Delta_R(-x) \nonumber
\end{align}
The definition of $\Delta_{R/A}$ is chosen such that $(\Box_x +m^2)\Delta_{R/A}(x-y)=\delta^{(4)}(x-y)$.

Furthermore, $P$ denotes the principal value of an integral, $\theta$ is the step function and
$\epsilon(x)=\theta(x)-\theta(-x)$ is the sign function.

\section{The Theorem of Epstein and Glaser}
\label{app:EG}

Epstein and Glaser consider vacuum expectation values ($l$-point functions) $F_{R/A}(p,q), p \in \R^{4l}, q \in \R^{4n}$ of time ordered products defined by retarded or advanced solutions. Obviously, they have retarded (advanced) support:
\begin{equation}
\label{eq:EGSupport1}
  \supp F_{R/A} \subset \biggl\{ (x,y) \in \R^{4(l+n)} | \{ y \} \subset \{x\}+\bar{V}_{\mp} \biggr\}.
\end{equation}
Here $\bar{V}_{\pm}$ denotes the closed forward/backward light cone. Furthermore, the difference of the retarded and the advanced product can be written as a sum over commutators, where on one side of the commutator only the interaction terms appear\footnote{This is easy to see in the modern notation $\mathcal{R}(e^{iA};e^{iB}) = S(B)^{-1} S(A+B)$, $\mathcal{A}(e^{iA};e^{iB}) = S(A+B) S(B)^{-1}$, see, e.g., \cite{HollandsWald}.}. Thus, if one computes the vacuum expectation value of this difference and inserts intermediary states between the factors of the commutator, the vacuum insertion vanishes. The insertions of projectors on the particle states then lead to the following support property of the Fourier transforms:
$\hat{F}_R(p,q) - \hat{F}_A(p,q) = 0$ for $q$ in
\begin{equation}
\label{eq:R}
   R_n= \left\{ q \in \R^{4n} | \left( \sum_{i \in I} q_i \right)^2 < 4 m^2 \text{ and } \neq m^2 \ \forall I \subset \{1, \dots , n \} \right\}
\end{equation}

Thus, one may apply the following
\begin{theorem}
\label{thm:EG}
If a pair of tempered distributions $F_{R/A} \in \schw'(\R^{4(l+n)})$ has the support (\ref{eq:EGSupport1}) and their Fourier transforms coincide for $q \in R_n$, then their Fourier transforms are tempered distributions in $p$ and infinitely differentiable in $q$ for all $q \in R_n$.
\end{theorem}

Thus, using a sequence $\{\check{G}_a(q)\}_a$ of test functions with support in a closed subset of $R_n$ that converges in $\mathcal{O}_C'(\R^{4n})$\footnote{Since $\hat{F}_{R/A}$ is a tempered distribution it can only have
polynomially bounded growth.} to $(2 \pi)^{2 n} \delta^{(4n)}$, one can carry out the adiabatic limit $a \to \infty$. The result is
independent of the sequence $\{ \check{G}_a \}_a$ and a tempered distribution in $p$. Of course it does not
matter if one uses the retarded or the advanced product.

\section{Proof of Lemma \ref{lemma:iep}}
\label{app:lemma}

We extend Lemma \ref{lemma:iep} to
\begin{lemma}
\label{lemma:2pv_pol} For $n\in \mathbb N$ and $l_r \in \mathbb N_0, r=0,\ldots,n$
\begin{equation}\label{equallemma1}
  \sum_{m=0}^n (-1)^m \prod_{r=0}^{m-1}\frac{(-\sum_{t=r+1}^m y_t)^{l_r}}{\sum_{t=r+1}^m y_t+i\epsilon}
  \cdot \delta_0^{l_m}\cdot \prod_{r=m+1}^{n}\frac{(\sum_{t=m+1}^r y_t)^{l_r}}{\sum_{t=m+1}^r y_t+i \epsilon}
  =P_{l_0,\ldots,l_n}(y_1,\ldots,y_n),
\end{equation}
where (with $a:=\sum_{t=0}^n l_t$) we have:
\begin{description}
  \item[I]: If $a<n$ then $P_{l_0,\ldots,l_n}=0$.
  \item[II]: If $a=n$ then $P_{l_0,\ldots,l_n}=1$.
  \item[III]: If $a>n$ then $P_{l_0,\ldots,l_n}(y_1,\ldots,y_n)$ is a homogeneous polynomial of degree $a-n$
  and further we have:
    \begin{description}
      \item[IIIa]: If $l_n=0$ the term with highest power in $y_n$ is $(-y_n)^{a-n}$.
      \item[IIIb]: If $n=2$ and $l_0=0$ the term with highest power in $y_1$ is $y_1^{a-n}$.
    \end{description}
\end{description}
\end{lemma}
\begin{proof}
The case $n=1$ is almost trivial, where we have of course $\frac{x_1}{x_1+i \epsilon}=1$ as a distribution. For
$n=2$ the cases where $(l_0,l_1,l_2)$ equals to (a permutation of) $(1,0,0), (1,1,0), (\ge 2,\ge 1, 0)$ or $(\ge
1, \ge 1, \ge 1)$ are easily checked. For the case $(l_0,l_1,l_2)=(0,0,0)$ we compute
\begin{multline*}
  \iep{y_1} \iep{y_1+y_2} - \iep{y_2} \iep{y_1} + \iep{y_1+y_2} \iep{y_2} \\
 = \iep{y_1+y_2} \frac{y_1+y_2}{y_1 y_2 + i(y_1+y_2)\epsilon} - \frac{1}{y_1 y_2 + i(y_1+y_2)\epsilon} = 0.
\end{multline*}
The remaining cases are permutations of $(b,0,0)$ with $b\ge2$. We show it here for $l_1=b$:
\begin{equation}\label{case0b0}
   \frac{(y_1)^{b-1}}{y_1+y_2+i\epsilon}  - \frac{(-y_2)^{b-1}}{y_1+y_2+i \epsilon} \\
    = \iep{y_1+y_2}(y_1+y_2)\sum_{k=0}^{b-2} y_1^{k} (-y_2)^{b-2-k}
    = \sum_{k=0}^{b-2} y_1^{k} (-y_2)^{b-2-k}.
\end{equation}
For the cases $l_0$ or $l_2 = b$ this can also be done and the parts IIIa and IIIb of the lemma are easily
checked explicitly.

Now we want to work by induction. For this, we assume $n\ge3$ and that the Lemma has been proven for all lower
orders. From the sum (\ref{equallemma1}) we split off the terms with $m=n$
\begin{equation*}
 (-1)^n \delta_0^{l_n}\prod_{r=1}^n\frac{(-\sum_{t=r+1}^n y_t)^{l_r}}{\sum_{t=r+1}^n y_t+i\epsilon}=:A
\end{equation*}
and with $m=n-1$
\begin{equation*}
 (-1)^{n-1} \delta_0^{l_{n-1}}\prod_{r=1}^{n-1}\frac{(-\sum_{t=r+1}^{n-1} y_t)^{l_r}}{\sum_{t=r+1}^{n-1} y_t+i\epsilon}
  \cdot \frac{y_n^{l_n}}{y_n+i \epsilon}=:B.
\end{equation*}
The remaining summands each have a factor
\begin{multline}\label{eqsplit}
  \frac{(\sum_{t=m+1}^{n-1} y_t)^{l_{n-1}}}{\sum_{t=m+1}^{n-1} y_t+i \epsilon}
  \frac{(\sum_{t=m+1}^n y_t)^{l_n}}{\sum_{t=m+1}^n y_t+i \epsilon}\\
  =\delta_0^{l_{n-1}} \frac{1}{\sum_{t=m+1}^{n-1} y_t} \frac{y_n^{l_n}}{y_n+i\epsilon}
   - \delta_0^{l_{n}} \frac{1}{\sum_{t=m+1}^{n} y_t}\frac{(-y_n)^{l_{n-1}}}{y_n+i\epsilon}
   + P_{0,l_{n-1},l_n}(\sum_{t=m+1}^{n-1} y_t, y_n),
\end{multline}
where we used the induction hypothesis for $n=2$. If we reinsert these terms into the remaining sum we can split
this into three parts, which we label according to the order in (\ref{eqsplit}) by $C,D$ and $E$. Now we can
combine $A+D$ to
\begin{multline}\label{aplusd}
  -\delta_0^{l_n}\frac {(-y_n)^{l_{n-1}}}{y_n+i \epsilon}
  \sum_{m=0}^{n-1} (-1)^m \prod_{r=0}^{m-1}\frac{(-\sum_{t=r+1}^m y^\prime_t)^{l^\prime_r}}{\sum_{t=r+1}^m y^\prime_t+i\epsilon}
  \cdot \delta_0^{l^\prime_m}\cdot \prod_{r=m+1}^{n-1}\frac{(\sum_{t=m+1}^r y^\prime_t)^{l^\prime_r}}{\sum_{t=m+1}^r y^\prime_t+i \epsilon}\\
  =-\delta_0^{l_n}\frac {(-y_n)^{l_{n-1}}}{y_n+i \epsilon} P_{l_0,\ldots,l_{n-2},0}(y_1,\ldots,y_{n-1}+y_n),
\end{multline}
with $l'_i = l_i$ and $y'_i = y_i$ for $i \leq n-2$ and $l'_{n-1} = 0$ and $x'_{n-1} = x_{n-1} + x_n$. The terms
$B+C$ give
\begin{multline}\label{bplusc}
  \delta_0^{l_{n-1}}\frac {y_n^{l_{n}}}{y_n+i \epsilon}
  \sum_{m=0}^{n-1} (-1)^m \prod_{r=0}^{m-1}\frac{(-\sum_{t=r+1}^m y_t)^{l_r}}{\sum_{t=r+1}^m y_t+i\epsilon}
  \cdot \delta_0^{l_m}\cdot \prod_{r=m+1}^{n-1}\frac{(\sum_{t=m+1}^r y_t)^{l_r}}{\sum_{t=m+1}^r y_t+i \epsilon}\\
  =\delta_0^{l_{n-1}}\frac {y_n^{l_{n}}}{y_n+i \epsilon} P_{l_0,\ldots,l_{n-2},0}(y_1,\ldots,y_{n-1}).
\end{multline}
Now we have a closer look at
\begin{equation*}
  E=\sum_{m=0}^{n-2} (-1)^m \prod_{r=0}^{m-1}\frac{(-\sum_{t=r+1}^m y_t)^{l_r}}{\sum_{t=r+1}^m y_t+i\epsilon}
  \cdot \delta_0^{l_m} \cdot \prod_{r=m+1}^{n-2}\frac{(\sum_{t=m+1}^r y_t)^{l_r}}{\sum_{t=m+1}^r y_t+i
  \epsilon}\cdot P_{0,l_{n-1},l_n}(\sum_{t=m+1}^{n-1} y_t, y_n).
\end{equation*}
The last polynomial gives $0$ if $l_{n-1}+l_n<2$. Otherwise, by IIIb, the term with highest power in
$\sum_{t=m+1}^{n-1}y_t$ from $P_{0,l_{n-1},l_n}$ is $(\sum_{t=m+1}^{n-1}y_t)^{l_{n-1}+l_n-2}$ and we can write
it as
\begin{equation*}
  P_{0,l_{n-1},l_n}(\sum_{t=m+1}^{n-1} y_t, y_n)=
  \sum_{\alpha=0}^{l_{n-1}+l_n-2}(\sum_{t=m+1}^{n-2}y_t)^{l_{n-1}+l_n-2-\alpha} \tilde P_\alpha (y_{n-1},y_n),
\end{equation*}
where $\tilde P_\alpha (y_{n-1},y_n)$ is a homogeneous polynomial of degree $\alpha$ and $\tilde P_0=1$. If
$l_n=0$ we can deduce from the explicit formula (\ref{case0b0}) that in each $\tilde P_\alpha (y_{n-1},y_n)$ we
have a term $(-y_n)^\alpha$. Now we pull in $E$ the sum over $\alpha$ to the front and for each summand use the
induction hypothesis for $n-2$ to get
\begin{equation}\label{equalE}
  E=\sum_{\alpha=0}^{l_{n-1}+l_n-2} P_{l_0,\ldots,l_{n-3},l_{n-2}+l_{n-1}+l_n-2-\alpha}(y_1,\ldots,y_{n-2})\cdot \tilde
  P_\alpha (y_{n-1},y_n).
\end{equation}
So $E$ is a homogeneous polynomial of degree $\sum_{r=0}^{n}l_r-2-(n-2)-\alpha+\alpha=\sum_{r=0}^{n}l_r-n$.

We have to check the following cases:
\begin{itemize}
  \item $l_{n-1}=l_n=0$: $E=0$ and
    \begin{equation*}
      A+D+B+C= \frac {1}{y_n+i \epsilon}
      \left[P_{l_0,\ldots,l_{n-2},0}(y_1,\ldots,y_{n-1})-P_{l_0,\ldots,l_{n-2},0}(y_1,\ldots,y_{n-1}+y_n)\right].
    \end{equation*}
    These polynomials are of degree $\sum_{r=0}^n l_r-(n-1)$ if this is greater or equal to $0$.
    If we expand the powers of $y_{n-1}+y_n$ of the second
    polynomial we see that terms with no factor $y_n$ vanish and from the remaining terms one factor is canceled by the
    prefactor. So the remaining expression is of degree $\sum_{r=0}^n l_r-n$ and I to IIIa are easily checked.
  \item $l_{n-1}=1,l_n=0$: $E=B+C=0$ and
    \begin{equation*}
    A+D=\frac {y_n}{y_n+i \epsilon} P_{l_0,\ldots,l_{n-2},0}(y_1,\ldots,y_{n-1}+y_n).
    \end{equation*}
    This is of degree $\sum_{r=0}^n l_r-(n-1)=\sum_{r=0}^n l_r-n$. Again, I to IIIa are easily checked.
  \item $l_{n-1}=0,l_n=1$: similar
  \item $l_{n-1}\ge2,l_n=0$: $B+C=0$. $A+D$ and $E$ both vanish if $\sum_{r=0}^n l_r-n<0$, so I is
  checked. Set $a^\prime:=\sum_{r=0}^{n-2} l_r$. To show II we assume $a^\prime+l_{n-1}-n=0$ from which
  follows $a^\prime-n\le-2$. So from (\ref{aplusd}) we see that the polynomial in $A+D$ vanishes. In $E$ only
  the term with $\alpha=0$ gives a contribution, which is $1$.\\
  Now we want to show III and IIIa: We have $a^\prime+l_{n-1}-n>0$ and see that both $A+D$ and $E$ are homogeneous
  polynomials of the right degree. We still have to show that not both are zero and they do not cancel each other.
  This is done if we
  show IIIa. For that, we have to look at the cases:
  \begin{enumerate}
    \item $a^\prime-n<-2$: $A+D=0$. The sum over $\alpha$ in $E$ only goes to $\alpha=a^\prime+l_{n-1}-n$ as for
    higher $\alpha$ the first polynomial in $E$ vanishes. The term with highest degree in $y_n$
    comes from $\alpha=a^\prime+l_{n-1}-n$ and is $(-y_n)^{a^\prime+l_{n-1}-n}$.
    \item $a^\prime-n=-2$: $A+D=0$ and in $E$ the term with highest $\alpha=l_{n-1}-2$ gives just
    $(-y_n)^{a^\prime+l_{n-1}-n}$. All other terms are of lower order in $y_n$.
    \item $a^\prime-n>-2$: The highest degree of $y_n$ in $E$ is $l_{n-1}-2<a^\prime +l_{n-1}-n$ whereas $A+D$
    gives a term $(-y_n)^{l_{n-1}-1}\cdot(-y_n)^{a^\prime-(n-1)}=(-y_n)^{a^\prime+l_{n-1}-n}$.
  \end{enumerate}
  \item $l_{n-1}=0,l_n\ge2$: similar
  \item $l_{n-1}\ge1,l_n\ge1$: $A+D=B+C=0$, only $E$ gives a contribution. I to III are again easily checked.
\end{itemize}
This completes the proof.
\end{proof}

\begin{acknowledgements}
We would like to thank Klaus Fredenhagen for valuable comments and discussions.  Financial support from the Graduiertenkolleg ``Zuk\"unftige Entwicklungen in der Teilchenphysik'' is gratefully acknowledged. Part of this work was done while J.~Z. visited the Dipartimento di Matematica of the Universit\`a di Roma ``La Sapienza'' with a grant of the research training network ``Quantum Spaces -- Noncommutative Geometry''. It is a pleasure to thank Sergio Doplicher for kind hospitality.
\end{acknowledgements}

\end{document}